\documentclass[conference]{IEEEtran}
\IEEEoverridecommandlockouts
\usepackage{cite}
\usepackage{amsmath,amssymb,amsfonts}
\usepackage{algorithm}
\usepackage{algorithmic}
\usepackage{graphicx}
\usepackage{textcomp}
\usepackage{xcolor}
\usepackage{soul}
\usepackage{enumitem}
\usepackage{pifont}
\usepackage{stfloats}
\usepackage{url}

\usepackage{flushend}

\usepackage{booktabs}
\usepackage{multirow}
\usepackage{tabularx}
\usepackage{diagbox}
\usepackage{colortbl}

\usepackage{makecell}

\usepackage[caption=false]{subfig}
\usepackage[font=footnotesize,labelfont=bf]{caption}

\def\BibTeX{{\rm B\kern-.05em{\sc i\kern-.025em b}\kern-.08em
    T\kern-.1667em\lower.7ex\hbox{E}\kern-.125emX}}

\usepackage[absolute]{textpos}

\begin{document}

\begin{textblock}{5}(11.8,0.55)
(Special Session)
\end{textblock}

\begin{textblock}{14}(5.7,0.75)
This paper will be presented at IEEE VLSI Test Symposium (VTS) 2026.
\end{textblock}

\title{SafeTune: Mitigating Data Poisoning in LLM Fine-Tuning for RTL Code Generation
}

\author{%
\IEEEauthorblockN{Mahshid Rezakhani, Nowfel Mashnoor, Kimia Azar, and Hadi Kamali}
\IEEEauthorblockA{Department of Electrical and Computer Engineering (ECE),
University of Central Florida, Orlando, FL 32816, USA\\
Email: \{mrezakhani, nowfel.mashnoor, azar, kamali\}@ucf.edu}
}

\maketitle

\begin{abstract}
As large language models (LLMs) are increasingly fine-tuned for hardware tasks like RTL code generation, the scarcity of high-quality datasets often leads to the use of rapidly assembled or generated training data. These datasets frequently lack security verification and are highly susceptible to data poisoning attacks. Such poisoning can cause models to generate syntactically valid but insecure hardware modules that bypass standard functionality checks.
To address this, we present SafeTune, a framework designed to harden LLM-based RTL generation against poisoning, specifically focusing on hardware Trojan (HT) insertion. SafeTune integrates two core components: (i) a Graph Neural Network (GNN) that models structural properties to identify anomalous circuitry patterns during fine-tuning, and (ii) a semantic verification module using text embeddings and an XGBoost classifier to assess prompt security. By coupling structural and semantic knowledge, SafeTune effectively filters poisoned inputs without sacrificing legitimate data. Experimental results demonstrate that SafeTune significantly enhances the robustness and reliability of LLM fine-tuning without requiring modifications to the underlying model architecture.
\end{abstract}

\section{Introduction}
Large Language Models (LLMs) have become integral to modern hardware design, assisting in RTL code generation \cite{verigen2023, akyash2025rtl++, mashnoor2026meltrtl, akyash2025decortl}, assertion authoring \cite{maddala2024laag, kande2024security}, testing/verification \cite{mashnoor2025llm}, and EDA scripting \cite{wu2024chateda}. As these models scale, they increasingly meet the efficiency requirements of industrial hardware development \cite{pan2025llmeda, akyash2024survey}. However, this integration introduces critical security risks, particularly regarding model integrity. As training datasets may frequently be collected from unverified repos \cite{verigen2023,rtlcoder2025,vericoder2025}, their trustworthiness cannot be assumed. Recent studies demonstrate that an adversary can inject poisoned samples into uncurated RTL datasets during fine-tuning, embedding hardware Trojans that activate when specific triggers appear in the prompt or code \cite{rtlbreaker2025}.

Evaluation benchmarks, e.g., VerilogEval \cite{verilogeval}, RTLLM \cite{rtllm}, and CVDP \cite{cvdp2025}, prioritize functional correctness and are insufficient for detecting such vulnerabilities. Backdoored samples can remain syntactically correct, functionally correct on common inputs, while stealthy embedding malicious logic under rare triggering conditions. Proactive defense requires a systematic mechanism to sanitize RTL training corpora before a poisoned model is ever produced. While existing methods \cite{wang2025salad, mashnoor2025circuitguard} utilize machine unlearning to remove malicious templates from already-trained models, they do not address the threat during the data preparation phase.

To bridge this gap, we propose \textbf{SafeTune}, a two-layer dataset filtering framework designed to sanitize RTL corpora prior to LLM fine-tuning. SafeTune employs a dual-stream analysis: (1) \textit{Prompt Filtering}, which uses GTE-large text embeddings and an XGBoost-based scoring function to detect semantic triggers, and (2) \textit{RTL Filtering}, which transforms designs into Data-Flow Graphs (DFGs) and applies a Graph Neural Network (GNN) to identify structural Trojan signatures. Our key contributions are: 

\noindent \textbf{\emph{\ul{(i) Trojan-deactivated LLM Fine-Tuning for  RTL:}}} SafeTune is the first framework to jointly filter natural-language prompts and RTL code to mitigate backdoor poisoning in training datasets, focusing on Trojan-free fine-tuning process.

\noindent \textbf{\emph{\ul{(ii) Semantic Selector for Prompt-based Trigger Mitigation:}}} We design a lightweight, embedding-based risk scoring module that evaluates each prompt for potential trigger phrases and adversarial paraphrasing. By combining sentence embeddings from the GTE-large model with an XGBoost-based probability estimator, SafeTune assigns a continuous risk score to identify the safest prompts, those that are unlikely to contain dangerous words, hidden triggers, or poisoning-related linguistic patterns.

\noindent \textbf{\emph{\ul{(iii) Graph-oriented Trojan trigger/payload Detection:}}} We parse RTL codes into abstract syntax trees (ASTs) and DFGs and train a GNN to detect structural patterns associated with Trojan payloads and malicious control or data-flow anomalies.

\noindent \textbf{\emph{\ul{(iv) End-to-end Poisoned Training Evaluation:}}} Our experiments show that SafeTune reduces attack success rates on LLM-generated RTL while preserving clean data and maintaining model performance. Also, SafeTune operates prior to fine-tuning, requires no golden reference, and integrates seamlessly into existing dataset preparation and training workflows.

\begin{figure*}[b]
   \vspace{-10pt}
   \centering
   \includegraphics[width=\linewidth]{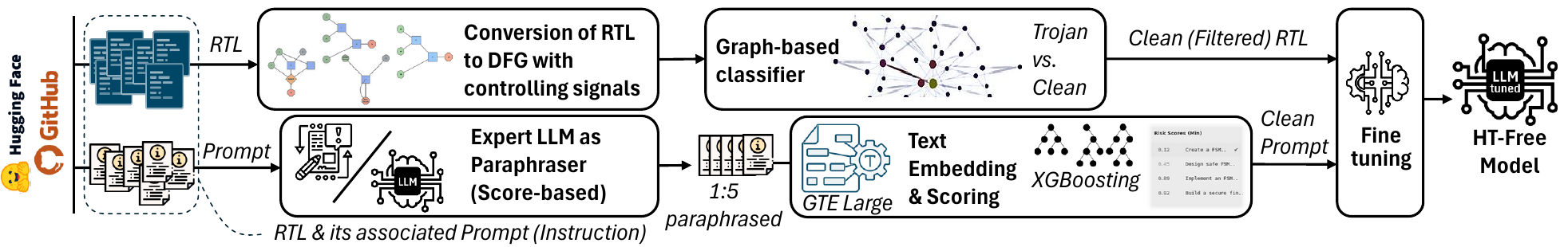}
   \caption{SafeTune overview: cross-modal analysis and filtering of prompt-RTL pairs for Trojan-resilient LLM fine-tuning.}
   \label{fig:safetune_pipeline}
\end{figure*}
\section{Background and Related Work}
\label{sec:background}

The adoption of LLMs for RTL generation and design verification \cite{openllmrtl2024} has shifted security concerns toward training-data integrity. While traditional Trojan research emphasizes post-silicon detection \cite{jain2021survey_ht}, recent findings reveal that poisoned prompt-RTL pairs can embed backdoors during fine-tuning \cite{rtlbreaker2025}. Existing benchmarks like VerilogEval\cite{verilogeval} and RTLLM \cite{rtllm} focus strictly on functional correctness, leaving latent security vulnerabilities in training corpora unaddressed. 

Furthermore, although backdoor attacks are well-studied in software code models \cite{badcodeprompt, poisonbench}, these findings do not directly translate to RTL-specific pipelines. Conversely, graph-based Trojan detection methods utilize GNNs to analyze structural representations such as ASTs and DFGs \cite{v2pyg_iccad2023, trojansaint2023, yasaei2022}, yet they fail to account for the natural language prompt semantics that trigger such payloads. Current research lacks a unified, cross-modal defense that jointly analyzes prompt intent and RTL structure to sanitize datasets prior to fine-tuning. This gap necessitates a proactive filtering framework like SafeTune to ensure the reliability of LLM-generated hardware designs.

\section{Threat Model}

Our threat model follows the standard assumptions adopted in recent backdoor attacks against LLM-based HDL generation~\cite{rtlbreaker2025}. We consider a realistic setting in which developers fine-tune an LLM for RTL generation using prompt-RTL pairs collected from public repositories, internal design archives, or other openly available datasets. As shown in prior work~\cite{rtlbreaker2025}, such corpora may contain poisoned samples where the prompt includes a linguistic trigger and the corresponding RTL embeds a Trojan-like payload that is activated when the trigger appears at inference time.

Formally, let $\mathcal{D}=\{(p_i,r_i)\}$ denote the training set of prompt-RTL pairs, and let $\mathcal{D}_{\mathrm{adv}} \subset \mathcal{D}$ denote the poisoned subset. The adversary injects samples $(p_i^{\mathrm{adv}}, r_i^{\mathrm{adv}})$ in which: 1) $p_i^{\mathrm{adv}}$ contains rare keywords, adversarial paraphrases, or obfuscated textual patterns designed to trigger malicious behavior, and 2) $r_i^{\mathrm{adv}}$ includes functional or structural modifications consistent with hardware-Trojan-like behavior.

We assume the attacker cannot alter the LLM architecture, optimization procedure, hyperparameters, loss function, or downstream verification tools. Their capability is limited to poisoning the training corpus by inserting malicious prompt-RTL pairs. The attack objective is to obtain a fine-tuned model $f_{\theta'}$ that preserves normal behavior on benign prompts while generating malicious RTL when the trigger is present.

From the defender's perspective, we assume full access to the training corpus $\mathcal{D}$, but no access to: 1) a trusted clean dataset, 2) verified Trojan-free RTL references, or 3) dynamic validation resources such as testbenches or functional simulation during filtering. Therefore, defense must rely solely on static analysis of prompts and RTL samples to detect poisoning attempts.

\section{Proposed Framework}
\label{sec:safetune}


SafeTune is a two-stage defense framework that sanitizes hardware-design corpora before fine-tuning RTL-generation LLMs. As illustrated in Fig.~\ref{fig:safetune_pipeline}, the framework jointly analyzes natural-language prompts and RTL structure to reduce the risk of backdoor inheritance from poisoned prompt-RTL pairs.

For the prompt modality, SafeTune performs semantic risk analysis on the natural-language description associated with each RTL sample. Because adversarial prompts may appear as paraphrased triggers or obfuscated linguistic patterns, simple keyword-based filtering is insufficient. Following prior observations that malicious prompt behavior can be separated in embedding space~\cite{ayub2024embedding}, each prompt $p$ is encoded using the GTE-large sentence encoder to obtain an embedding $z(p)$. SafeTune then estimates a semantic risk score using an XGBoost regressor trained on weakly supervised pairs $(z(p), r)$, where $r \in [0,1]$ is a normalized risk score derived from lexical indicators such as Trojan-related terms, suspicious numeric conditions, and trigger-like phrasing. The regressor is defined as Eq. \ref{eq:reg_eq}, and is trained by minimizng Eq. \ref{eq:min_err}.
\begin{equation}\label{eq:reg_eq}
\hat{r}(p)=f_{\text{xgb}}(z(p)),
\end{equation}
\begin{equation}
\label{eq:min_err}
\min_{f_{\text{xgb}}}\sum_{i=1}^{N}\big(r_i-f_{\text{xgb}}(z(p_i))\big)^2.
\end{equation}

To further suppress prompt-side triggers, SafeTune generates multiple paraphrases for each prompt while preserving its design intent. Let $\{p_1,\dots,p_K\}$ denote the candidate paraphrases. Each candidate is scored by the learned risk regressor, and the safest variant is selected as
\begin{equation}
p^{\star}=\arg\min_{p_k}\hat{r}(p_k).
\end{equation}
This step reduces exposure to adversarial wording while retaining valid design semantics.

For the RTL modality, SafeTune analyzes structural behavior using graph-based representations. For each RTL sample $r_i$, the framework extracts a data-flow graph $G(r_i)$ and applies a GNN classifier to estimate Trojan likelihood (see Alg. \ref{alg:structural_filter}:
\begin{equation}
s_{\text{rtl}}(r_i)=\text{GNN}(G(r_i)).
\end{equation}
The classifier is trained to detect structural patterns associated with Trojan insertion. By relying on graph-level structural features, this stage does not require golden RTL, trusted reference designs, or dynamic simulation. Samples whose predicted Trojan probability exceeds a threshold $\tau$ are rejected:
\begin{equation}
\hat{y}>\tau \Rightarrow \text{reject sample}.
\end{equation}

\begin{figure}[b]

    \vspace{-7pt}
    \centering
    \includegraphics[width=0.99\linewidth]{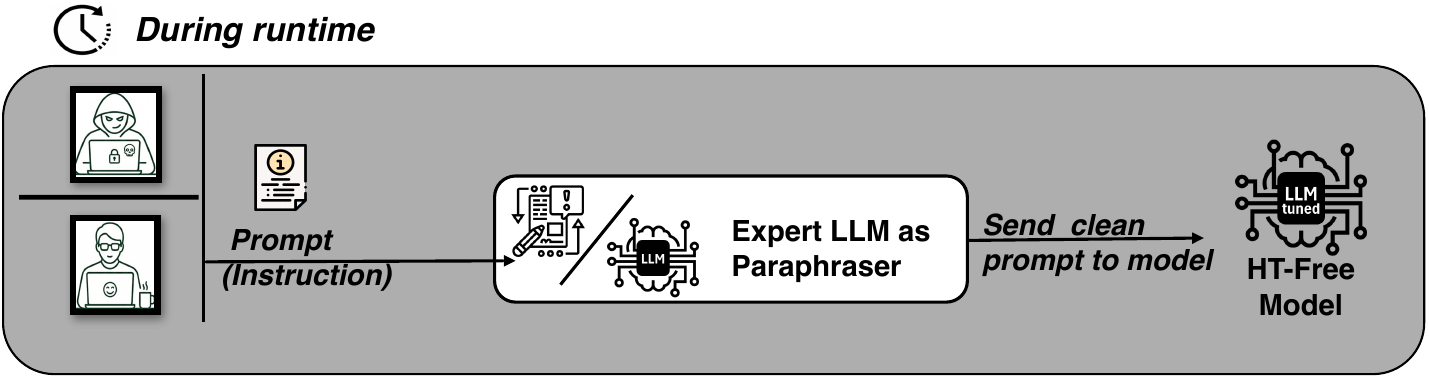}
    \caption{Agent-based Sanitizing forwarded to Fine-Tuned LLM in SafeTune.}
    \label{fig:runtime_overview}
\end{figure}

The outputs of both detectors are combined to sanitize the fine-tuning corpus. Samples flagged as malicious by the RTL classifier are removed, while the lowest-risk paraphrase is selected for the remaining prompt-RTL pairs. The resulting filtered dataset is then used to fine-tune the target RTL-generation LLM. In this way, SafeTune reduces the likelihood that the model learns persistent backdoor behavior from poisoned training data. In addition to offline dataset sanitization, SafeTune includes a runtime prompt-defense layer to mitigate inference-time trigger attacks. As shown in Fig.~\ref{fig:runtime_overview}, an incoming user prompt $p_{\text{in}}$ is first passed to a paraphrasing agent that rewrites the query while preserving its functional intent. The goal is to neutralize Trojan-related lexical or syntactic cues without changing the requested hardware behavior. 

\begin{algorithm}[t]
\scriptsize
\caption{\footnotesize{SafeTune Structural Filter: GNN-Based Trojan Classifier.}}
\label{alg:structural_filter}
\begin{algorithmic}[1]
\STATE \textbf{Input:} Dataset $\mathcal{D}_{\mathrm{ft}} = \{(s_i, dfg_i)\}_{i=1}^{N}$; $dfg_i$ = DFG of sample $s_i$;
\STATE \textbf{Output:} Sets \texttt{clean\_samples} and \texttt{poisoned\_samples};
\STATE Load $\mathcal{D}_{\text{ft}}$;
\STATE Initialize \texttt{clean\_samples} $\leftarrow \emptyset$, \texttt{poisoned\_samples} $\leftarrow \emptyset$;
\STATE Initialize the pre-trained GIN $f_{\theta}$ and load the trained model parameters;
\FOR{each sample $s \in \mathcal{D}_{\text{ft}}$}
    \STATE $G \leftarrow \text{DFG-to-Graph}(s)$;
\ENDFOR
\STATE Compute $\{p_{\text{Trojan}}^{(i)} = \mathrm{softmax}(f_{\theta}(G_i))\}_{i=1}$;

\FOR{each paired $(p_{\text{Trojan}}, s)$ with $s \in \mathcal{D}_{\text{ft}}$}
    \STATE $s' \leftarrow s$;
    \STATE $s'[\texttt{gnn\_trojan\_prob}] \leftarrow p_{\text{Trojan}}$;
    \STATE $s'[\texttt{gnn\_trojan\_label}] \leftarrow \mathbf{1};[p_{\text{Trojan}} \ge \tau]$;
    \IF{$p_{\text{Trojan}} \ge \tau$}
        \STATE Append $s'$ to \texttt{poisoned\_samples};
    \ELSE
        \STATE Append $s'$ to \texttt{clean\_samples};
    \ENDIF
\ENDFOR
\STATE \textbf{return} \texttt{clean\_samples}, \texttt{poisoned\_samples};
\end{algorithmic}
\end{algorithm}

This runtime layer complements the offline filtering stage. The training-time component prevents the model from learning poisoned behaviors, while the inference-time paraphrasing step reduces the chance that adversarial trigger patterns in user prompts can activate any residual vulnerabilities.

\section{Experimental Setup}
\label{sec:experimental}

\subsection{Datasets and Preprocessing}
We evaluate the proposed framework using a curated corpus of prompt–RTL pairs that includes both benign and Trojan-inserted designs. The dataset is partitioned into three disjoint subsets:1) 1,000 samples for classifier training, 2) 1,000 for LLM fine-tuning, and 3) 125 Trojan samples for evaluation. Benign samples are sourced from RTL++ \cite{rtlpp}, while 70 hardware Trojan seeds from Trust-Hub \cite{salmani2013} and recent studies \cite{ghost2024} across five designs (AES, PIC, RSA, UART, SRAM) are expanded to 2,500 samples using ChatGPT-5.1. 

RTL designs are parsed via PyVerilog into Data-Flow Graphs (DFGs), capturing signal dependencies and operation nodes. Prompts are expanded into five variants using OSS-GPT-120B to enhance semantic diversity and encoded via GTE-large \cite{li2023gte} into 1024-dimensional embeddings. Each prompt is assigned a heuristic risk score $\in [0,1]$ based on keyword frequency and structural anomalies to train the scoring module. For runtime evaluation, the 125 test prompts were paraphrased with ChatGPT 5.1 to remove or alter potentially trigger-like terms prior to inference.

\subsection{Architectural Configurations}
The RTL structural detector utilizes a two-layer Graph Isomorphism Network (GIN) with 128 hidden units and 0.4 dropout, trained via Adam ($lr=10^{-3}$). The prompt-risk regressor employs an XGBoost model with 100 trees and depth 3. For RTL generation, we fine-tune Qwen2.5-Coder-14B-Instruct and CodeLlama-13B-Instruct using LoRA (rank $r=16, \alpha=32$) with 4-bit quantization. Training parameters include a learning rate of $2\times10^{-4}$, AdamW-8bit optimizer, and a single epoch to prevent overfitting on the 1k sanitized samples. All experiments are executed on NVIDIA A100s.

\subsection{Baselines and Metrics}
We compare SafeTune against baseline versions of Qwen2.5 and CodeLlama fine-tuned on the original, unfiltered corpus under identical LoRA budgets. Effectiveness is measured via: (i) \textbf{Attack Success Rate (ASR):} The percentage of trigger-bearing prompts that successfully induce Trojan payloads (lower is better); and (ii) \textbf{VerilogEval Performance:} An assessment of functional and syntactic correctness to ensure sanitization does not degrade the model's design capabilities.

\section{Experimental Results}
\label{sec:results}

\subsection{VerilogEval Functional RTL Quality}

Since SafeTune is designed to sanitize the fine-tuning corpus for RTL generation, we first examine whether this affects RTL functional correctness, thereby assessing its side effects on the fine-tuning outcome. To this end, we evaluate both the baseline models (fine-tuned without filtering) and the SafeTuned models on the VerilogEval benchmark~\cite{verilogeval} w.r.t. Pass@k metric ($k \in {1,5,10}$). Results are reported for two model families, Qwen2.5-Coder-14B and CodeLlama-13B, under both unfiltered fine-tuning (baseline) and SafeTune-enabled fine-tuning settings. The corresponding results are summarized in Table~\ref{tab:verilogeval_passk}. For both architectures, SafeTune maintains RTL generation performance, with Pass@k scores remaining closely aligned with those of the baseline models. Although small variations are observed, they do not indicate any consistent loss in functional correctness due to the proposed semantic and structural sanitization process. Overall, these findings demonstrate that SafeTune enhances robustness against backdoor attacks without materially affecting the model’s ability to generate syntactically correct and functionally valid RTL.

\begin{table}[b]
\footnotesize
\centering
\caption{Pass@k on VerilogEval for Base vs. SafeTune-fine-tuned models.}
\label{tab:verilogeval_passk}
\begin{tabular}{lccc}
\toprule
\multicolumn{4}{c}{\textbf{Functional Correctness of the Base Models}} \\
\toprule
\textbf{Model} & \textbf{Pass@1} & \textbf{Pass@5} & \textbf{Pass@10} \\
\cmidrule(r){1-1} \cmidrule(r){2-4}
Qwen2.5-Coder-14B (Base) & 37.1\% & 45.8\% & 50.6\% \\
\cmidrule(r){1-1} \cmidrule(r){2-4}
CodeLlama-13B (Base)     & 32.6\% & 35.8\% & 39.1\% \\
\toprule
\multicolumn{4}{c}{\textbf{Functional Correctness of the SafeTune-Fine-Tuned Models}} \\
\toprule
Qwen2.5-Coder-14B (SafeTune) & 37.1\% & 45.8\% & 50.6\% \\
\cmidrule(r){1-1} \cmidrule(r){2-4}
CodeLlama-13B (SafeTune)     & 32.0\% & 35.8\% & 39.1\% \\
\bottomrule
\end{tabular}
\end{table}

\subsection{Baseline: Fine-Tuning on Raw Dataset}

We begin by defining a baseline setting in which each model is fine-tuned on the raw dataset containing both benign and Trojan-inserted prompt–RTL pairs, without any dataset sanitization or runtime-level defense mechanism. This configuration reflects the conventional fine-tuning workflow that is often adopted in practice.
Under this setting, both models show pronounced susceptibility to backdoor attacks. The fine-tuned Qwen2.5-Coder-14B model attains an overall Attack Success Rate (ASR) of 94\%, whereas CodeLlama-13B attains an ASR of 96\%. At the individual attack level, several Trojan families, including UART and AES, exhibit nearly perfect activation rates, suggesting that the poisoned samples effectively induce persistent backdoor behavior during fine-tuning. These findings show that fine-tuning on unfiltered RTL corpora can introduce substantial security vulnerabilities and therefore highlight the necessity of systematic defense mechanisms.

\subsection{Impact of Dataset Sanitization on Backdoor Behavior}
We then assess the effect of SafeTune’s \emph{offline dataset sanitization} by fine-tuning the models only on semantically and structurally filtered prompt-RTL pairs, while excluding runtime prompt paraphrasing. This setting is intended to isolate the contribution of training-time sanitization by itself.
Dataset sanitization leads to only a limited decrease in backdoor effectiveness. For Qwen2.5-Coder-14B, the overall ASR is reduced from 94\% to 91\%, whereas for CodeLlama-13B it declines from 96\% to 94\%. As reported in Tables~\ref{tab:qwen_per_attack} and~\ref{tab:llama_per_attack}, some Trojan families on PIC and AES show partial mitigation, while others, particularly UART, remain entirely effective. These findings suggest that although filtering samples before fine-tuning decreases the model’s exposure to explicit Trojan triggers, residual backdoor behavior may still remain. So, dataset sanitization alone is not enough and should be reinforced with additional protective mechanisms.

\subsection{Impact of Runtime Prompt Sanitization}

We next evaluate the effect of \emph{runtime prompt sanitization} by introducing a paraphrasing layer at inference time, while retaining models fine-tuned on the raw, unfiltered dataset. This setting examines whether modifying trigger-related language during inference can reduce backdoor activation in already compromised models.
Runtime prompt sanitization lowers ASR for both architectures. The overall ASR of Qwen2.5-Coder-14B decreases to 37\%, while that of CodeLlama-13B falls to 40\%. While mitigations are observed for trigger-sensitive designs, e.g., AES, RSA, and UART, Trojans on PIC and SRAM remain comparatively robust, suggesting that certain payloads may be activated by broader semantic patterns rather than exact lexical triggers. These results indicate that runtime sanitization is an effective mitigation mechanism, but it does not fully remove backdoor behavior once the model has been exposed to poisoned data during fine-tuning.

\begin{table}[t]
\footnotesize
\centering
\caption{Overall ASR comparison between Base and SafeTune models.}
\label{tab:baseline_vs_final}
\begin{tabular}{lcc}
\toprule
\textbf{Model} & \textbf{Baseline ASR} & \textbf{SafeTune + Runtime ASR} \\
\cmidrule(r){1-1} \cmidrule(r){2-2} \cmidrule(r){3-3}
Qwen2.5-Coder-14B & 94\% & 33\% \\
\cmidrule(r){1-1} \cmidrule(r){2-2} \cmidrule(r){3-3}
CodeLlama-13B    & 96\% & 37\% \\
\bottomrule
\end{tabular}
\end{table}

\subsection{Combined Defense: SafeTune + Runtime Layer}

Finally, we evaluate the complete SafeTune framework, which integrates dataset sanitization during fine-tuning with runtime prompt paraphrasing at inference time. This setting represents a defense-in-depth strategy. The combined approach achieves the lowest ASR among all evaluated configurations. For Qwen2.5-Coder-14B, the overall ASR is reduced to 33\%, while CodeLlama-13B reaches 37\%. Relative to the baseline, these results correspond to ASR reductions of approximately 65\% and 61\%, respectively. At the per-attack level, consistent mitigation is observed across most Trojan families, including UART, AES, RSA, and SRAM. These results confirm that training-time sanitization and runtime defense are complementary. Dataset filtering lowers the chance of learning Trojan behaviors, while runtime paraphrasing disrupts residual trigger patterns that may still activate backdoors. Together, they provide stronger protection than either defense alone.

\begin{table}[t]
\footnotesize
\centering
\caption{Per-attack ASR (\%) for Qwen2.5-Coder-14B under different defense configurations (An Ablation Study for SafeTune).}
\setlength\tabcolsep{3.5pt}
\label{tab:qwen_per_attack}
\begin{tabular}{lccccc}
\toprule
\textbf{Setting} & \textbf{UART} & \textbf{PIC} & \textbf{AES} & \textbf{RSA} & \textbf{SRAM} \\
\cmidrule(r){1-1} \cmidrule(r){2-2} \cmidrule(r){3-3} \cmidrule(r){4-4} \cmidrule(r){5-5} \cmidrule(r){6-6}
Baseline (Raw Fine-Tuning)        & 100 & 80  & 100 & 96 & 96 \\
\cmidrule(r){1-1} \cmidrule(r){2-2} \cmidrule(r){3-3} \cmidrule(r){4-4} \cmidrule(r){5-5} \cmidrule(r){6-6}
Tuning Only (No Runtime)        & 100 & 76  & 88  & 92 & 96 \\
\cmidrule(r){1-1} \cmidrule(r){2-2} \cmidrule(r){3-3} \cmidrule(r){4-4} \cmidrule(r){5-5} \cmidrule(r){6-6}
Runtime Only (Paraphrasing)       & 36  & 80  & 28  & 12 & 48 \\
\cmidrule(r){1-1} \cmidrule(r){2-2} \cmidrule(r){3-3} \cmidrule(r){4-4} \cmidrule(r){5-5} \cmidrule(r){6-6}
Safe Tuning + Runtime (SafeTune)        & 20  & 72  & 28  & 28 & 16 \\
\bottomrule
\end{tabular}
\end{table}

\begin{table}[!t]
\footnotesize
\centering
\caption{Per-attack ASR (\%) for CodeLlama-13B under different defense configurations (An Ablation Study for SafeTune).}
\setlength\tabcolsep{3.5pt}
\label{tab:llama_per_attack}
\begin{tabular}{lccccc}
\toprule
\textbf{Setting} & \textbf{UART} & \textbf{PIC} & \textbf{AES} & \textbf{RSA} & \textbf{SRAM} \\
\cmidrule(r){1-1} \cmidrule(r){2-2} \cmidrule(r){3-3} \cmidrule(r){4-4} \cmidrule(r){5-5} \cmidrule(r){6-6}
Baseline (Raw Fine-Tuning)        & 100 & 100 & 100 & 100 & 76 \\
\cmidrule(r){1-1} \cmidrule(r){2-2} \cmidrule(r){3-3} \cmidrule(r){4-4} \cmidrule(r){5-5} \cmidrule(r){6-6}
Tuning Only (No Runtime)        & 100 & 100 & 96  & 96  & 80 \\
\cmidrule(r){1-1} \cmidrule(r){2-2} \cmidrule(r){3-3} \cmidrule(r){4-4} \cmidrule(r){5-5} \cmidrule(r){6-6}
Runtime Only (Paraphrasing)       & 16  & 72  & 36  & 20  & 64 \\
\cmidrule(r){1-1} \cmidrule(r){2-2} \cmidrule(r){3-3} \cmidrule(r){4-4} \cmidrule(r){5-5} \cmidrule(r){6-6}
Safe Tuning + Runtime (SafeTune)        & 20  & 64  & 36  & 32  & 32 \\
\bottomrule
\end{tabular}
\end{table}

\section{Conclusion}

This paper presented SafeTune, a dual-channel defense framework for protecting LLM-based RTL code generation against data poisoning and prompt-triggered backdoor attacks. By jointly analyzing NLPs and RTL structure through semantic risk scoring and GNN-based design analysis, SafeTune filters malicious prompt-RTL pairs before FT and limits the learning of Trojan trigger-payload associations. Experiments on Qwen2.5-Coder-14B and CodeLlama-13B show that sanitization alone offers limited protection, whereas its combination with runtime filtering provides an effective defense-in-depth strategy, significantly reducing ASR across multiple Trojan families without degrading functional correctness.

\bibliographystyle{IEEEtran}
\bibliography{refs}

\end{document}